\documentclass[aps,prl,twocolumn,preprintnumbers,superscriptaddress,
              showpacs,nofootinbib]{revtex4-1}
\newcommand{\PRE}[1]{}       

\usepackage{bm} \usepackage{epsfig}

\usepackage{slashed}

\def\beq{\begin{eqnarray}}
\def\eeq{\end{eqnarray}}
\def\bea{\begin{eqnarray}}
\def\eea{\end{eqnarray}}

\newcommand{\kev}{\text{keV}}

\newcommand{\gev}{\text{GeV}}

\newcommand{\pb}{\text{pb}}

\newcommand{\cm}{\text{cm}}

\newcommand{\s}{\text{s}}

\newcommand{\eqref}[1]{Eq.~(\ref{#1})}

\newcommand{\tableref}[1]{Table~\ref{table:#1}}

\newcommand{\gsim}{\lower.7ex\hbox{$\;\stackrel{\textstyle>}{\sim}\;$}}
\newcommand{\lsim}{\lower.7ex\hbox{$\;\stackrel{\textstyle<}{\sim}\;$}}

\newcommand{\sigmaSI}{\sigma_{\rm SI}}

\def\gev{\, {\rm GeV}}

\def\kev{\, {\rm keV}}

\newcommand{\ssection}[1]{{\bf #1\ }}

\begin{document}

\preprint{UCI-TR-2011-29, UH511-1182-11}

\title{ \PRE{\vspace*{1.5in}}
New Constraints on Isospin-Violating Dark Matter
\PRE{\vspace*{0.3in}} }

\author{Jason Kumar}
\affiliation{Department of Physics and Astronomy, University of
Hawaii, Honolulu, HI 96822, USA
\PRE{\vspace*{.2in}}
}

\author{David Sanford}
\affiliation{Department of Physics and Astronomy, University of
California, Irvine, CA 92697, USA
\PRE{\vspace*{.2in}}
}

\author{Louis E.~Strigari}
\affiliation{
Kavli Institute for Particle Astrophysics and Cosmology, Stanford
University, Stanford, CA 94305 USA
\PRE{\vspace*{.4in}}
}

\date{December 2011}

\begin{abstract}
\PRE{\vspace*{.3in}}
We derive bounds on the dark matter annihilation cross-section for
low-mass ($5-20~\gev$) dark matter annihilating primarily to up or
down quarks, using the Fermi-LAT bound on gamma-rays from Milky Way
satellites.  For models in which dark matter-Standard Model
interactions are mediated by particular contact operators, we show
that these bounds can be directly translated into bounds on the dark
matter-proton scattering cross-section.  For isospin-violating dark
matter, these constraints are tight enough to begin to constrain the
parameter-space consistent with experimental signals of low-mass dark
matter.  We discuss possible models that can evade these bounds.
\end{abstract}


\maketitle

\ssection{Introduction.}  There has been great interest in low-mass
dark matter ($m_X \sim 5-20~\gev$) as a possible way to explain
potential signals from the DAMA~\cite{Bernabei:2010mq},
CoGeNT~\cite{Aalseth:2011wp} and CRESST~\cite{Angloher:2011uu}
experiments.  However, this data seems to be in tension with exclusion
bounds from XENON10~\cite{Angle:2011th},
XENON100~\cite{Aprile:2011hi},
CDMS~\cite{arXiv:1010.4290,Ahmed:2010wy} and
SIMPLE~\cite{Felizardo:2011uw}.  Of these, the constraints from the
xenon-based experiments appear to be the tightest.  Attention has thus
turned to new models which can potentially reconcile this data.  One
focus has been on isospin-violating dark matter
(IVDM)~\cite{Kurylov:2003ra,Giuliani:2005my,Fitzpatrick:2010em,Chang:2010yk,Feng:2011vu,arXiv:1105.3734,arXiv:1109.4639},
in which dark matter interactions between protons and neutrons are
different.  Although further reconciliation is required for complete
consistency~\cite{Kopp:2011yr}, destructive interference between dark
matter couplings to protons and to neutrons can potentially relieve
the tension between the CoGeNT, DAMA and XENON10/100 data sets.  IVDM
may thus play an important role in understanding the low-mass data.

A consequence of destructive interference is that the coupling of dark
matter to up- and down-quarks must be relatively large for a given
value of the dark matter-nucleus spin-independent scattering cross-section.
This translates into larger annihilation and production cross-sections.
Indirect detection and collider searches, where
destructive interference plays no role, can thus provide
stronger constraints on such dark matter models.
In particular, the large annihilation
cross-sections can enhance photon signals from the decays of hadrons
produced in dark matter annihilation to up- and down-quarks. Because
of their sensitivity to dark matter with mass in the range $\sim 5-20$
GeV, gamma-ray searches of dwarf spheroidal
galaxies~\cite{collaboration:2011wa,GeringerSameth:2011iw,Garde:2011hd}
can constrain these models.

In this Letter, we determine the constraints on the dark matter
annihilation cross-section arising from gamma-ray searches of dwarf
spheroidals, assuming dark matter annihilates entirely to up or down
quarks.  We then apply these constraints to IVDM models.  We will see
that if isospin-violation is the key feature in alleviating the
tension between the DAMA, CoGeNT and XENON10/100 data sets, then
models with purely contact interactions are tightly constrained by
Fermi-LAT data.  We also discuss how constraints on IVDM models weaken
when dark matter-quark interactions occur through a light mediator.

\ssection{Bounds from dwarf spheroidals.}
In~\cite{GeringerSameth:2011iw}, bounds on the dark matter
annihilation cross-section were determined from a combined analysis of
gamma-rays from dwarf spheroidals.  This analysis compared the
observed number of high-energy photons arriving directly from the
dwarf spheroidals to the observed background slightly off-axis from
the dwarf spheroidal.  These bounds were expressed in terms of the
quantity $\Phi_{PP}$, defined as
\bea
\Phi_{PP} &=& {\langle \sigma_A v \rangle \over 8\pi m_X^2}
\int_{E_{thr}}^{m_X}\sum_f  B_f {dN_f \over dE} dE ,
\label{BoundEqn}
\eea
where $B_f$ is the branching ratio for dark matter to annihilate to
channel $f$.  $dN_f / dE$ is the photon spectrum for each annihilation
channel and $E_{thr}$ is the detector's threshold photon energy.
Using $E_{thr}=1~\gev$,~\cite{GeringerSameth:2011iw} found
the 95\% CL bound
\bea
\Phi_{PP} \leq 5.0_{-4.5}^{+4.3} \times 10^{-30}~\cm^3~\s^{-1}~\gev^{-2},
\label{PhiBound}
\eea
where the asymmetric uncertainties
are 95\% CL systematic errors~\cite{collaboration:2011wa}, resulting
from uncertainty in the mass density profile of the various
satellites.
95\% CL exclusion contours on the dark matter annihilation cross-section can thus be
determined from eqs.~\ref{BoundEqn},\ref{PhiBound} for any $m_X$, using
the integrated photon distribution from each
annihilation channel.

\begin{figure}[tb]
\includegraphics[width=0.95\columnwidth]{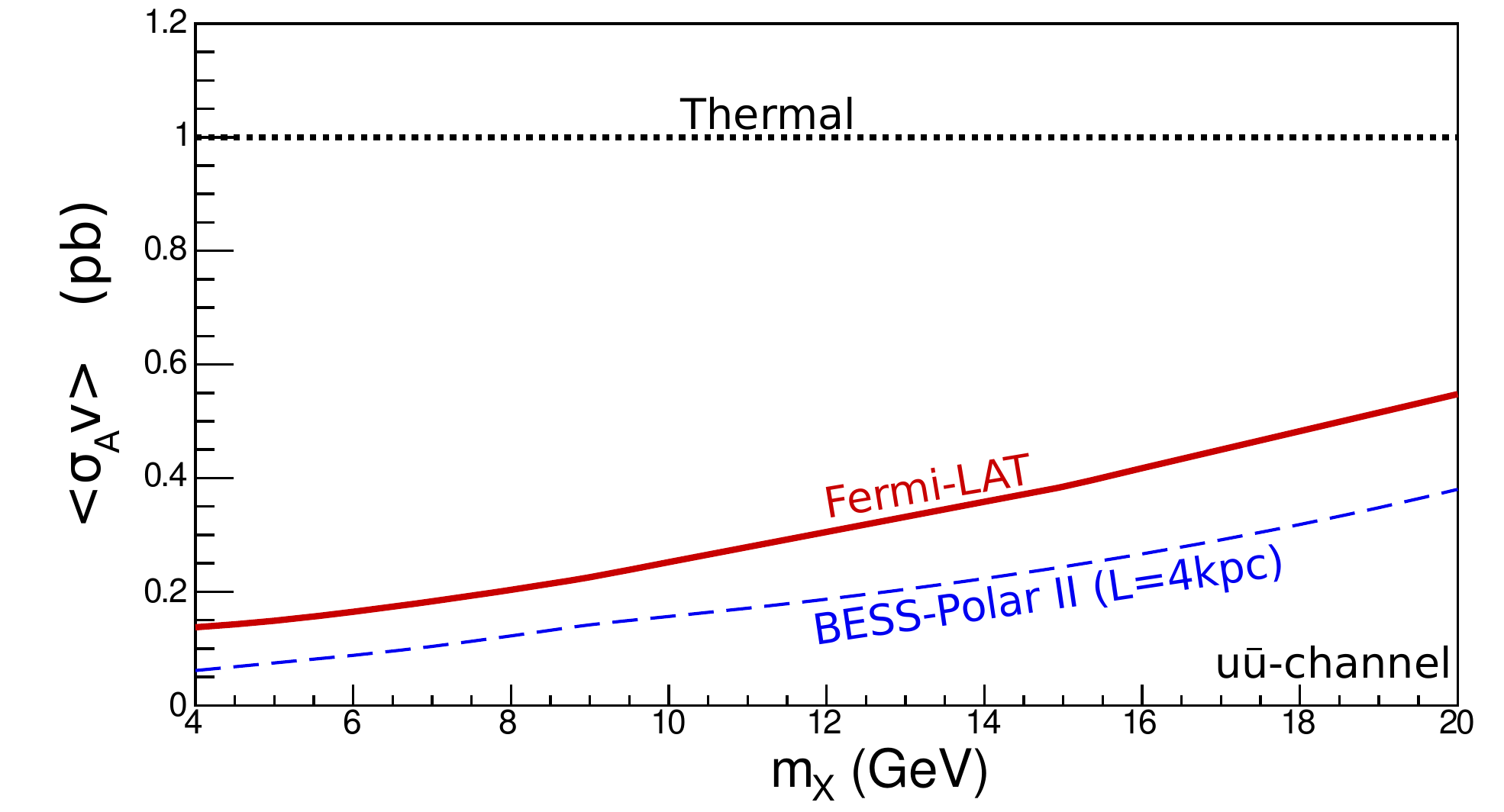}
\vspace*{-.1in}
\caption{\label{fig:sigmaSIp} 95\% CL exclusion bound (see text) from
  Fermi in the $(m_X, \langle \sigma_Av \rangle )$ plane, assuming
  dark matter annihilation to $u \bar u$.  The results for
  annihilation to the $d \bar d $ channel are visually identical.
  Also shown is the exclusion contour from BESS-Polar II for the same
  channel (assuming a diffusion halo size of 4 kpc).  The dotted line
  indicates $1~\pb$, which is approximately the total annihilation
  cross-section at freeze-out for dark matter produced thermally in
  the early universe.}
\end{figure}

For annihilation to up or down quarks, we generated the photon
spectrum using Pythia 6.409~\cite{hep-ph/0603175}, run on the
computing cluster at the Hawaii Open Supercomputing Center.  The up
and down quark channels produce more photons than the $\tau$-channel,
resulting in tighter bounds.  The bound for the $u \bar u$ channel is
plotted in figure~\ref{fig:sigmaSIp}; the bound for the $d \bar d$
channel is slightly tighter than that for the $u \bar u$ channel, but the
difference is never more than a percent for the given mass range.  For
these bounds, we have assumed the central value of $\Phi_{PP} \leq
5.0$, applicable for the satellite mass density profiles found
in~\cite{collaboration:2011wa}; allowing this to vary within 95\%
systematic error bars could weaken this bound by a factor of up to
$\sim 1.9$, or strengthen it by a factor of up to $\sim 10$.

Dark matter annihilation to up- and down-quarks can also be
constrained by limits on the cosmic ray anti-proton flux from
experiments such as BESS-Polar II~\cite{arXiv:1110.4376} and
Pamela~\cite{arXiv:0912.4510,arXiv:1007.5253}.  Of these two
experiments, BESS-Polar II reports tighter bounds, which we have
plotted in figure~\ref{fig:sigmaSIp} (light blue dashed curve,
assuming a diffusion halo size of 4 kpc).  The reported BESS-Polar II
bound on dark matter annihilation to up- or down-quarks is tighter
than the Fermi-LAT bound, given the astrophysics assumptions
underlying both analyses.  However, there are significant systematic
astrophysical uncertainties which can weaken the anti-proton flux
limits by up to a factor $\sim 50$~\cite{Evoli:2011id}. These include
uncertainties regarding the background and the dark matter density
profile of the galaxy, as well as uncertainties in the cosmic ray
propagation model.  Solar modulation can also sensitively affect
bounds on dark matter annihilation arising from anti-proton flux
measurements.  Depending on these systematic uncertainties, the
constraints on dark matter annihilation from the anti-proton flux may
be weaker than those from gamma-ray searches of dwarf spheroidals.

The anti-proton flux uncertainties which we have discussed above are
largely independent of the systematic uncertainties for gamma-rays
from dwarf spheroidals, which derive primarily from the dark matter
distribution within the spheroidal.  These two indirect detection
methods are thus quite complementary.

\ssection{IVDM.}  We now apply these constraints to models of
isospin-violating dark matter.  We will consider the case in which
dark matter interactions with Standard Model particles (both
scattering and annihilation) are mediated by effective contact
operators.  The most general such effective operator can be written in
terms of dark matter and quark bilinears, each of which transforms as
either a scalar, pseudoscalar, vector, pseudovector, tensor or
pseudotensor.  For such interactions, the dark matter-proton
spin-independent scattering cross-section ($\sigmaSI^p$) can be
unambiguously related to the dark matter annihilation cross-section.

We will be interested in IVDM models which couple only to up and down
quarks, since these provide conservative bounds.  Couplings to all
other quarks are isospin-invariant, and would result in a larger dark
matter annihilation cross-section for fixed $\sigmaSI^p$.  We will
only consider operators which yield spin-independent,
velocity-independent scattering matrix elements.  Furthermore, one
will only obtain relevant bounds from Fermi data if the annihilation
matrix element is $s$-wave.  Given these choices, there are only
three effective operators to consider:
\bea
{\cal  O}_{D} &=& C_{D}^q {1 \over M_*^2}\bar X \gamma^\mu X \bar q
\gamma_\mu q
\nonumber\\
{\cal O}_{C} &=& C_{C}^q {1 \over M_*} X^* X
\bar q q
\nonumber\\
{\cal O}_{R} &=& C_{R}^q {1 \over M_*} X^2 \bar q q
\label{opeqn}
\eea
where ${\cal O}_{D,C,R}$ are possible couplings if the dark matter
($X$) is a Dirac fermion, complex scalar or real scalar,
respectively.\footnote{Majorana fermions are not considered because
  there is no corresponding operator producing both $s$-wave
  annihilation and velocity-independent scattering.  Also, it is
  useful to note that the most popular Majorana candidate, the
  neutralino in supersymmetry, is at least marginally inconsistent with
  the observations from CoGeNT, DAMA, and
  SIMPLE~\cite{Kuflik:2010ah}.}  Here, $C_{D,C,R}^q$ are dimensionless
couplings, $M_*$ is an overall energy scale and $q=u,d$.  The WIMPless
model~\cite{arXiv:0803.4196,arXiv:0806.3746,arXiv:0808.2318} exhibited
in~\cite{Feng:2011vu} as an example of IVDM was a real scalar coupling
through ${\cal O}_R$.

The $\sigmaSI^{p,n}$ can then be written as
\bea
\sigmaSI^{p,n(D)} &=&   {\mu_p^2 f_{p,n}^2 \over \pi M_*^4 }
\nonumber\\
\sigmaSI^{p,n(C)} &=&   {\mu_p^2 f_{p,n}^2 \over 4\pi M_*^2 m_X^2 }
\nonumber\\
\sigmaSI^{p,n(R)} &=&   {\mu_p^2 f_{p,n}^2 \over \pi M_*^2 m_X^2 }
\label{sigmaSIpeqn}
\eea
where $\mu_p$ is the dark matter-proton reduced mass and the $f_{p,n}$
are coefficients parameterizing the coupling of dark
matter to protons and neutrons, respectively.  For $f_n / f_p \sim -0.7$,
the constraints from XENON10/100 are maximally suppressed because of almost exact
cancelation between proton and neutron interactions in the xenon
target~\cite{Chang:2010yk,Feng:2011vu}.  Moreover, for this choice of
$f_n / f_p$, the regions
favored by DAMA and CoGeNT are brought into alignment.

The $f_{p,n}$ are related to the coefficients $C^{u,d}$ by
\bea
f_p &=& C^u B_u^p + C^d B_d^p
\nonumber\\
f_n &=& C^u B_u^n + C^d B_d^n
\label{fnfpeqn}
\eea
where $B_{u,d}^{p,n}$ are the integrated nuclear form-factors~\cite{astro-ph/0110225}.  When
the dark matter is a scalar it couples through operator ${\cal O}_{C,R}$ to the scalar quark
current, and the form factors are roughly given by $B_u^{p(scl.)}= B_d^{n(scl.)}\sim 6$ and
$B_u^{n(scl.)} = B_d^{p(scl.)} \sim 4$.
When the dark matter is a Dirac fermion it couples through operator ${\cal O}_D$ to the vector quark
current, and the appropriate form factors are $B_u^{p(vec.)}= B_d^{n(vec.)}= 2$
and $B_u^{n(vec.)} = B_d^{p(vec.)} = 1$.

The annihilation cross-section arising from each effective operator can be
written as
\bea
\langle \sigma_A^D v \rangle &=&  {3m_X^2 \over \pi M_*^4}(|C^u|^2 + |C^d|^2)
\nonumber\\
\langle \sigma_A^C v \rangle &=& {3 \over 4\pi M_*^2}(|C^u|^2 + |C^d|^2)
\nonumber\\
\langle \sigma_A^R v \rangle &=& {3 \over \pi M_*^2}(|C^u|^2 + |C^d|^2)
\label{sigmaAveqn}
\eea
If the dark matter is a Dirac fermion or complex scalar, we will assume
that the dark matter particle and anti-particle number densities in Milky Way
satellites are equal.

\begin{figure}[tb]
\includegraphics[width=0.95\columnwidth]{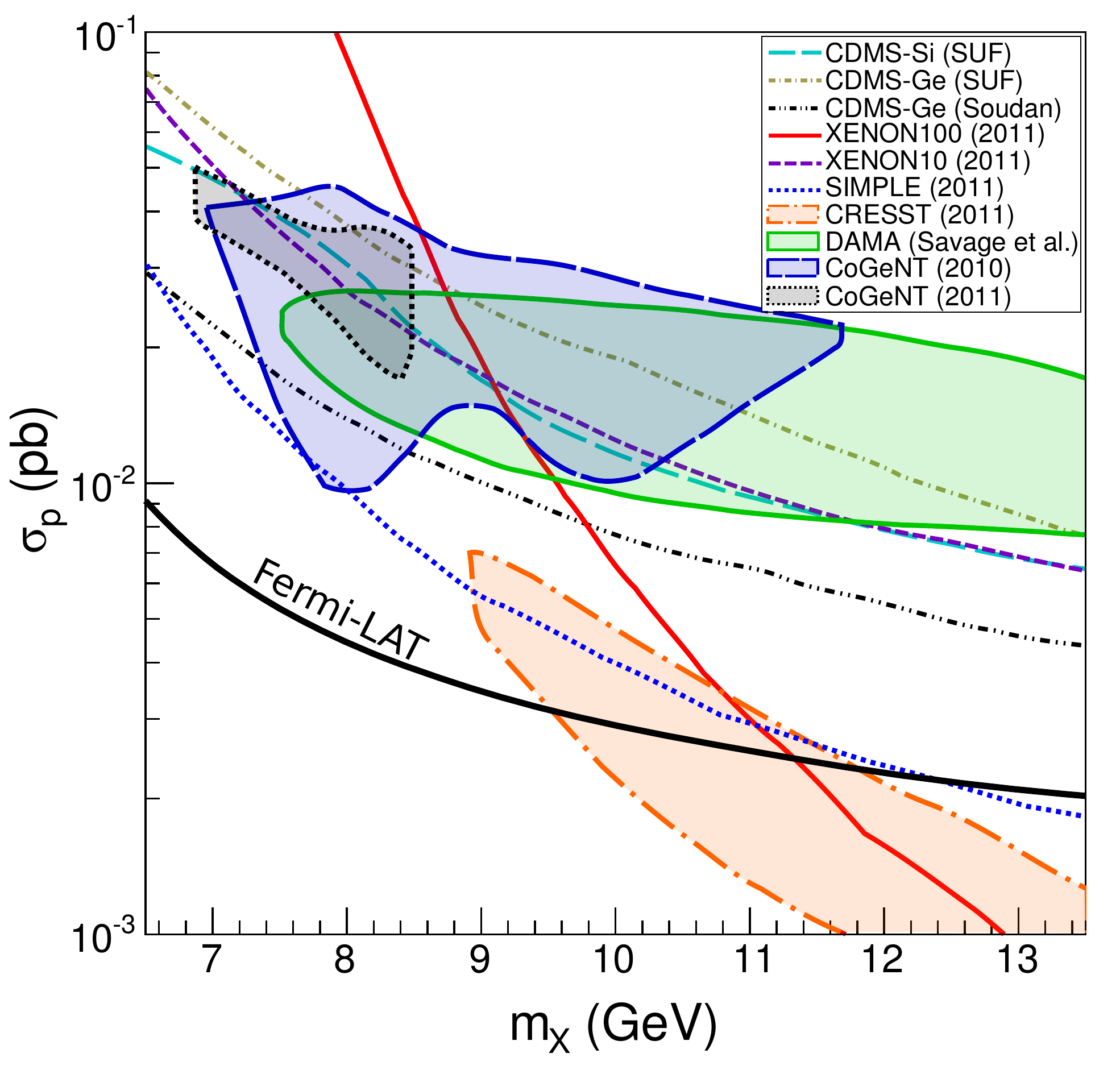}
\vspace*{-.1in}
\caption{\label{fig:sigmaann} Favored regions and exclusion contours
  in the $(m_X, \sigma_p)$ plane for IVDM with $f_n / f_p = - 0.7$.
  The solid black line is the bound from this analysis at 95\% CL,
  assuming central values for the satellite density profile.}
\end{figure}

For any choice of $f_n / f_p$ and choice of the interaction operator, one can use
eqns.~\ref{sigmaSIpeqn},\ref{fnfpeqn},\ref{sigmaAveqn} to obtain constraints on $\sigmaSI^p$ from
the bounds in figure~\ref{fig:sigmaSIp}.
These constraints are shown in figure~\ref{fig:sigmaann} for $f_n / f_p \sim -0.7$,
assuming dark matter interacting through ${\cal O}_C$ (if dark matter
couples through ${\cal O}_R$ or ${\cal O}_D$, the bounds would be tighter by a factor of $2$ or
$\sim 4$, respectively).
Also shown are the signal regions of DAMA~\cite{Bernabei:2010mq} ($3\sigma$, assuming no
channeling~\cite{arXiv:0808.3607,arXiv:1006.0972}), CoGeNT~\cite{Aalseth:2011wp} (90\% CL)
and CRESST~\cite{Angloher:2011uu} ($2\sigma$), along with
90\% CL exclusion contours from XENON10~\cite{Angle:2011th}, XENON100~\cite{Aprile:2011hi},
CDMS~\cite{arXiv:1010.4290,Ahmed:2010wy} and SIMPLE~\cite{Felizardo:2011uw}.
As we can see from this figure, the CoGeNT region of interest would be excluded by
the Fermi data, under the assumptions made here.

The CoGeNT experiment has recently reported
a preliminary analysis indicating that their experiment may have more surface area contamination
than originally thought~\cite{CollarTaup2011,arXiv:1110.5338}.  The effect of this correction would be to shift
the CoGeNT region of interest to slightly larger mass and slightly smaller $\sigmaSI^p$, though the
magnitude of the shift can only be determined by a complete analysis from the CoGeNT collaboration.
This shift would serve to weaken constraints from Milky Way satellites.

\ssection{Uncertainties from particle physics and astrophysics.}
There are a several ways in which the bounds from Fermi data can be
weakened.  For example, if the effective interaction operator is
generated by the exchange of a mediating particle with mass $m_\phi
\sim 1~\gev$, then the momentum transfer during scattering
interactions at any direct detection experiment (${\cal O}(10~\kev)$)
would be much smaller than the mediator mass~\cite{arXiv:1108.4661}.
As a result, scattering interactions would still be described by an
effective contact operator.  However, the annihilation cross-section
would receive an additional suppression of $(m_\phi / 2m_X )^4 \sim
10^{-4}-10^{-5}$.  In this case, models with $\sigmaSI^p$ which could
match the CoGeNT data would be unconstrained by limits from dwarf
spheroidals by $\sim 3-4$ orders of magnitude.

In addition, bounds from Fermi would be insignificant if dark matter
interactions were mediated by an operator yielding
velocity-independent spin-independent scattering, but $p$-wave
suppressed annihilation.  An example of such an operator would be
$(1/M_*^2)\bar X X \bar q q$, in the case where the dark matter is a
fermion.  Another example would be $(1/M_*^2) X^* \partial_\mu X \bar
q \gamma^\mu q$, if the dark matter is a complex scalar.  Dark matter
coupling through these operators would have an annihilation cross-section
which is suppressed by $v^2$; the constraints from Fermi on $\sigmaSI^p$
would thus be suppressed by more than 6 orders of magnitude.  These
models could potentially explain the
low-mass data of DAMA, CoGeNT and XENON10/100, but would be
unconstrained by the Fermi-LAT dwarf spheroidal search.

Both $p$-wave suppressed annihilation and light mediators would
similarly weaken the bounds from anti-proton flux.  In either case,
the constraints from the anti-proton flux are suppressed by the same
factor as the gamma-ray bounds, as discussed above.  Such models
matching the CoGeNT, DAMA and XENON10/100 data would be unconstrained
by bounds from BESS-Polar II.

The dark matter-proton scattering cross-section needed to match
low-mass direct detection data can be shifted by other particle
physics or astrophysics uncertainties, such as uncertainties in the
nuclear form factor or in the dark matter velocity distribution near
the earth.  To normalize with all of the observations, here we have
used the standard halo model to calculate the WIMP velocity
distribution~(e.g.~\cite{arXiv:1005.0579}); variations from this model
may affect the results presented, especially for high mass
targets~\cite{arXiv:1010.4300}.

\ssection{Complementary searches.}  There are interesting constraints
on the annihilation of low-mass dark matter from WMAP-7
data~\cite{arXiv:1103.2766,arXiv:1106.1528}.  These constraints arise
from the effect on standard recombination of the photons produced by
dark matter annihilation.  In the mass range we consider (assuming $u$/$d$-channel annihilation),
the bounds which we obtain from dwarf spheroidals
are a factor of $\sim 5$ tighter than those arising from WMAP-7 data.  Note
there are some systematic uncertainties in the WMAP-7 bounds regarding the effect of proton production
on heating of the cosmic medium~\cite{arXiv:1106.1528}.  Again, these uncertainties
are different from and independent of the uncertainties in dwarf spheroidal searches.
Planck data is expected to significantly improve these CMB bounds.

There have been several recent studies of constraints on low-mass dark
matter arising from the neutrino flux yielded by annihilation in the
sun~\cite{Fitzpatrick:2010em,arXiv:0808.2464,arXiv:0808.4151,arXiv:0908.1768,arXiv:1103.3270,arXiv:1106.4044}.
But neutrino detectors cannot constrain dark matter which annihilates
primarily to first generation quarks.  The final state quarks
hadronize, and these light hadrons typically lose energy and stop in
the sun before decaying.  They thus produce a neutrino spectrum far
too soft to be distinguished from the atmospheric neutrino background.

The Fermi-LAT gamma-ray bounds are also complementary to recent constraints from collider monojet
searches~\cite{Feng:2005gj,Goodman:2010ku,Rajaraman:2011wf,Goodman:2010yf,Goodman:2011jq}
on effective contact operators coupling dark matter to quarks.
However, applying previous results to the case considered here is
difficult, since they have typically assumed a dark matter coupling to
all quarks.  In particular, bounds in previous studies depend strongly
on the existence of heavy quark operators, and are weakened
considerably when only couplings to first generation quarks are
present.

To address this issue, we provide simple bounds for $\sigmaSI^p$ given
couplings only to up and down quarks with $f_n/f_p = -0.7$.  These
bounds are based on an ATLAS monojet search using 1~fb$^{-1}$ of
data~\cite{Aad:2011xw}.  We present bounds for two possible selection
cuts on $\slashed{E}_T$ and on the $p_T$ of the leading jet: $p_T >
120~\gev$, $\slashed{E}_T > 120~\gev$ and $p_T > 350~\gev$,
$\slashed{E}_T > 300~\gev$.  In addition to these selection cuts, the
ATLAS study also used an intermediate cut of $\slashed{E}_T >
250~\gev$, $p_T > 220~\gev$.  The bounds provided by this intermediate cut
lie between those of the other two, and are not shown.
For each choice of selection cuts, the ATLAS
collaboration determined a bound on the $pp \rightarrow XXj$
cross-section attributable to new physics, based on a comparison of
the number of observed events to the expected number of background
events.

For IVDM models, we generated the production cross-section for each of
the effective operators in eqn.~\ref{opeqn} using MadGraph/MadEvent
5.1.3~\cite{Alwall:2011uj}.  The selection cuts were imposed at the
parton level.  According to the analysis
of~\cite{Goodman:2010ku,Goodman:2011jq,WillEff}, approximately 40\% of
the events satisfying the cuts at parton level will satisfy the cuts
once hadronization and detector effects are included.  Accounting for
this 40\% selection efficiency, we determined upper bounds on
$C_{C,R}^{u,d} / M_*$ for the scalar case and $C_D^{u,d} / M_*^2$ for
the Dirac case.  These translate directly into the upper bounds on
$\sigmaSI^p$ given in \tableref{collider} for the two sets of missing
energy and $p_T$ cuts.  Equivalent Tevatron bounds from CDF monojet
searches~\cite{CDFmonojet} are weaker than the ATLAS bounds (assuming
the $p_T > 120~\gev$, $\slashed{E}_T > 120~\gev$ selection cut) by a
factor of $\sim 4$.  If dark matter is a real or complex scalar the
monojet search sensitivity is significantly weaker than the
sensitivity of Fermi-LAT and direct detection experiments.  Only in
the case of Dirac dark matter are collider bounds comparable to the
Fermi-LAT constraint.

\begin{table}
\begin{tabular}{ccc}
\begin{tabular}{|r|r|r|r|}
    \hline
    $m_X$ & ${\cal O}_{D}$ & ${\cal O}_{C}$ & ${\cal O}_{R}$    \\
    \hline
    4 & 0.00285 & 90.6 & 181 \\
    7 & 0.00320 & 35.4 & 70.3 \\
    10 & 0.00357 & 18.9 & 37.6 \\
    15 & 0.00370 & 9.1 & 18.2 \\
    20 & 0.00380 & 5.4 & 10.9 \\
    \hline
    \end{tabular}
    & \qquad\qquad &
    \begin{tabular}{|r|r|r|r|}
    \hline
    $m_X$ & ${\cal O}_{D}$ & ${\cal O}_{C}$ & ${\cal O}_{R}$    \\
    \hline
    4  &  0.00079  &  10.8  &  21.6 \\
    7  &  0.00092  &   4.2  &  8.50 \\
    10 &  0.00097  &   2.3  &  4.51 \\
    15  &  0.00106  &  1.1  &  2.13 \\
    20  &  0.00107  &  0.62  &  1.24 \\

    \hline
    \end{tabular} \\
\small{(a) $p_T > 120~\gev$} & & \small{(b) $p_T > 350~\gev$} \\
\small{\ \ \ \ $\slashed{E}_T > 120~\gev$} & & \small{\ \ \ \ $\slashed{E}_T > 300~\gev$}
\end{tabular}
\caption{Upper bounds on $\sigmaSI^p$ in pb from ATLAS monojet
  searches, assuming the listed cuts on the leading jet $p_T$ and on
  the missing transverse energy.  Masses are listed in GeV.  The
  columns correspond to Dirac fermions, complex scalars, and real
  scalars respectively.}
\label{table:collider}
\end{table}

However, in the case of light mediators, collider bounds become
significantly weaker.  Assuming that the $\bar q q \rightarrow \bar X
X$ process proceeds through the $s$-channel, the collider
cross-section scales as $(p_\phi^2 - m_\phi^2)^{-2}$.  For light
mediators we have $m_\phi \ll |p_\phi| > 2m_X$.  This leads to a
collider cross-section which is suppressed by at least $(2m_X)^{-4}$, instead of
$m_\phi^{-4}$.  The corresponding rescaling of the direct detection
bound relative to the contact operator case is at least $(2m_X /
m_\phi)^4$, weakening the bound
significantly~\cite{Goodman:2011jq}.  If $m_\phi \sim 1~\gev$, the bound
on $\sigmaSI^p$ would be raised by a factor of $\sim 10^4 -
10^5$, placing relevant direct detection cross-sections for any
species of dark matter well out of range of collider studies even at
the LHC, at least in the conservative case of couplings only to light
quarks and through low-mass mediators.\footnote{Note that for light
  mediators, the CDF monojet search~\cite{CDFmonojet} can place bounds
  which are tighter than the ATLAS search by a factor $\sim
  10$~\cite{Friedland:2011za}.  These bounds still do not constrain
  the IVDM models discussed here.}

\ssection{Conclusions.} We have shown that Fermi-LAT gamma-ray
searches of Milky Way satellites can constrain the
cross-section for low-mass dark matter to annihilate to up or
down-quarks.  These constraints are tighter than for models with
annihilation primarily to b-quarks or $\tau$-leptons, due to the
larger number of photons arising from annihilation to first-generation
quarks.  This is especially interesting in the case of
isospin-violating dark matter, where destructive interference between
up and down quark couplings imply an enhancement in the annihilation
cross-section for models which can match the low-mass data of DAMA,
CoGeNT and XENON10/100.  In particular, IVDM models whose effective
contact interactions yield $s$-wave annihilation are tightly
constrained.  Nevertheless, there are IVDM models which can evade these
Fermi-LAT bounds and other constraints from indirect detection and
collider searches.  These IVDM models involve either low-mass
mediators, or effective interaction operators which yield $p$-wave
suppressed annihilation.

It is worth noting that microscopic dark matter models may involve
interactions mediated by more than one effective operator.  To
understand constraints on such models from gamma-ray searches, one
would need to know how the coefficients of the various interaction
effective operators are correlated in any particular model.

In the case where dark matter annihilates to first generation quarks,
the sensitivity of indirect search strategies depends on the
spectrum of first generation quarks.  The photon spectrum arising from
the hadronization and decay of first generation quarks is peaked at
low values of $E_\gamma / m_X$.  Future searches using Milky Way
satellites would thus greatly benefit in any reduction in the
gamma-ray analysis threshold, $E_{thr}$.

\ssection{Acknowledgments.} We gratefully acknowledge S.~Koushiappas
and W.~Shepherd for useful discussions, and Jonathan L.~Feng for
initial collaboration on this project.  We also thank the Hawaii Open
Supercomputing Center, and D.~Hafner for facilitating our use of HOSC.
JK and LES are grateful to Aspen Center for Physics, where part of
this work was done, for their hospitality.  JK is supported by DOE
grant DE-FG02-04ER41291.  DS is supported by NSF grant PHY–0970173 and
a UC Irvine Graduate Dean’s Dissertation Fellowship.




\end{document}